\documentclass[amsmath,amssymb,times,11pt]{article}
 
\newtheorem{thm}{Theorem}[section] 
\newtheorem{lm}[thm]{Lemma}

\newcommand{\qed}{\mbox{\rule{1.6mm}{4.3mm}}}

\begin{document} 

\title{Optimal Encryption of Quantum Bits}

\author{P. Oscar Boykin and Vwani Roychowdhury
\thanks{E--mail addresses of the authors are, respectively:
\tt \{boykin, vwani\}@ee.ucla.edu.} 
\\ \small Electrical Engineering Department \\ \small UCLA
\\ \small Los Angeles, CA 90095}

\date{March 14, 2000}

\maketitle

\begin{abstract}
We characterize the complete set of protocols that may
be used to securely encrypt $n$ quantum bits using
secret and random classical bits.  
In addition to the
application of such quantum encryption protocols to quantum data security, our
framework allows for generalizations of many classical cryptographic 
protocols to quantum data. 
We show that the 
encrypted state gives no information without the secret 
classical data, and that $2n$ random classical bits are
the minimum necessary for informationally secure quantum encryption.
Moreover, the quantum operations are shown to have a surprising
structure in a canonical inner product space.
This quantum encryption protocol is a generalization 
of the classical one time pad concept.  A connection is made between
quantum encryption and quantum teleportation\cite{teleport}, and this allows
for a new proof of optimality of teleportation.
\end{abstract}

\section{Introduction}
We consider informationally secure encryption protocols, where 
any potential eavesdropper, Eve, will have no information about the original
quantum state,  even if she manages to steal or intercept the entire encrypted 
quantum data. 
This scenario is very different from the well-known scheme of quantum
cryptography, which  in the usual sense\cite{BB84,B92} is really
a secure expansion of an existing classical key, using a quantum channel and
a pre-selected set of quantum states.  The resulting secure bits might 
then be used for an encryption algorithm on classical 
data.  But suppose one is concerned with securing quantum data, as is the
case considered in this paper.  Extending ideas from QKD (such as testing bits 
in conjugate bases), one might show that given the test is passed, the quantum 
bits are also secure.  However, this case is ill-suited to data security 
as opposed to communication security. For the tasks targeted in the paper,
we need a method to make sure that even if the eavesdropper takes the 
quantum data, she will still learn nothing about the quantum information. 
In this case, the eavesdropper may not care about passing any tests, 
and may remove the qubits and replace them with junk.

We provide a simple method to get informationally secure encryption of
any quantum state using a classical secret key.  This could have several interesting
applications.  For example, if we imagine a scenario where good quantum memories are 
expensive, one might rent quantum storage.  Security in such a public-storage model 
would be a high priority. We assume the user cannot store quantum data herself, but
can store classical data.
Methods of using trusted centers for quantum cryptography have been
developed\cite{BHM}.
Our method would allow a user to encrypt her
quantum data using a classical key and allow a potentially malicious center
to store the data, and yet she would know that the center could learn nothing
about her stored quantum data. Additionally, the untrusted center
could act as a quantum communication provider.
Several other applications which involve adaptations 
of classical cryptographic protocols, such as quantum secret sharing using 
classical key, are outlined later in the paper.

\section{Classical Informationally Secure Encryption}
If $M$ is the random variable for the message, and 
$C$ is the random variable for the ciphertext (i.e., output
of the encryption process), then Shannon defined informationally secure 
cryptography in the
following way\cite{Shannon_Crypto}:
\begin{equation}
I(M;C)=H(C)-H(C|M)=0\ .
\end{equation}
The above relationship implies $p(c|m)=p(c)$, i.e., that the ciphertext, $c$,
is independent of the message, $m$.
Since one must be able to recover the message from the ciphertext
given the key, one must also satisfy $I(M;C|K)=H(M)$.  Hence, the secrecy 
condition combined with the recoverability condition imply that $H(K)\ge H(M)$
and $H(C)\ge H(M)$ 
for informationally secure cryptography.

An example of informationally secure cryptography is the one time
pad\cite{one_time_pad}.  
The message $m$ is compressed to it's entropy, and then a full-entropy 
random string of length $H(M)$ is chosen and called $k$. Then, the ciphertext 
is $c=m\oplus k$.  Given $c$, one knows nothing of $m$,
but given $c$ and $k$, one has $m$ exactly.

This same one time pad approach may be applied in the quantum case.

\section{Encryption of Quantum Data}
Alice has a quantum state that she intends either to send to Bob, or to 
store in a quantum memory for later use.  Eve may intercept the state 
during transmission or may access the quantum memory.  Alice
wants to make sure that {\em even if} Eve receives the entire state, she learns
nothing. Toward this end, any encryption algorithm must be a unitary 
operation, or more specifically a set of unitary operations which may 
be chosen with some distribution. It must be unitary because one must 
be able to undo the encryption, and any quantum operation that is 
reversible is unitary\cite{Preskill_Notes}.

The most general scheme is to have a set of $M$ operations, 
$\{U_k\}$, $k=1, \ldots, M$, where each element $U_k$ is a $2^n \times 2^n$
unitary matrix.  This set of unitary operations is assumed to be 
known to all, but the classical key, $k$, which specifies the $U_k$ that
is applied to the $n$-bit quantum state,  is secret.
The key is chosen with some probability $p_k$ and the input quantum
state is encrypted by applying the corresponding unitary operation $U_k$.
In the decryption stage, $U_k^\dagger$ is applied to the quantum state to
retrieve the original state.

The input state, $\rho$, is called the message state, and the output 
state, $\rho_c$, is called the cipher-state. The protocol is secure if for
every input state, $\rho$, the output state, $\rho_c$,  is the totally mixed state: 
\begin{equation}
\label{qcrypt_cond}
\rho_c=\sum_k p(k)U_k\rho U_k^{\dagger}=\frac{1}{2^n}I \ .
\end{equation}
The reason that $\rho_c$ must be the totally mixed state is two fold.  First,
for security all inputs must be mapped to the same output density matrix
(because $\rho_c$ must be independent of the input).
Second, the output must be the totally mixed state because the totally 
mixed state is clearly mapped to itself by all encryption sets. 

To see that this is secure, we note that Eve could prepare an n-bit totally
mixed state on her own.  Since two processes that output the same
density matrices are indistinguishable\cite{Peres}, anything that 
can be learned from $\rho_c$ can also be learned from the totally mixed state.  

The {\em design criterion is to find such a distribution of unitary operations
$\{p_k,U_k\}$ that will map all inputs to the totally mixed state.}  
A construction of such a map is given next.

\section{A Quantum One Time Pad}
\label{onetime_sec}
  The algorithm is simple: for each qubit, Alice and Bob share
two random secret bits.  We assume these bits are shared in advance.
If the first bit is $0$ she does nothing, else
she applies $\sigma_z$ to the qubit.  If the second bit is $0$ she does
nothing, else she applies $\sigma_x$.  Now she sends the qubit to Bob. She 
continues this protocol for the rest of the bits.

We now show that this quantum one time pad protocol is secure. First note that
this bit-wise protocol can be expressed in terms of our general quantum encryption 
setup by choosing $p_k=1/2^{2n}$ and $U_k=X^\alpha Z^\beta$ 
($\alpha, \beta \in \{0,1\}^n$), where $\displaystyle X^\alpha=
\bigotimes_{i=1}^n\sigma_x^{\alpha(i)}$ and $\displaystyle Z^\beta=
\bigotimes_{i=1}^n\sigma_z^{\beta(i)}$. Thus $X^\alpha$ corresponds to
applying $\sigma_x$ to the bits in positions given by the $n$-bit 
string $\alpha$, and similarly for $Z^\beta$.  Next, define the inner product
of two matrices, $M_1$ and $M_2$, as $Tr(M_1 M_2^\dagger)$. If the set of all $2^n\times 2^n$ 
matrices is seen as an inner product space (with respect to the preceding inner product),
then one can easily verify that the set of $2^{2n}$ unitary matrices $\{X^\alpha Z^\beta\}$ forms an 
orthonormal basis.   Expanding any message state, $\rho$, in this $X^\alpha Z^\beta$ basis gives:
\begin{equation}
\rho=\sum_{\alpha,\beta}a_{\alpha,\beta}X^\alpha Z^\beta \ ,
\end{equation}
where $a_{\alpha,\beta}=Tr(\rho Z^\beta X^\alpha)/2^n$.  Using this
formalism, it is clear that the given choice of $p_k$ and $U_k$ satisfies 
eqn. (\ref{qcrypt_cond}), and hence the underlying protocol is secure: 
\begin{eqnarray}\label{xz_proof}
\sum_k p(k)U_k\rho U_k^{\dagger}&=&
\frac{1}{2^{2n}}\sum_{\gamma,\delta}X^\gamma Z^\delta\rho Z^\delta X^\gamma 
\nonumber \\
&=&\frac{1}{2^{2n}}\sum_{\alpha,\beta}a_{\alpha,\beta}\sum_{\gamma,\delta}
X^\gamma Z^\delta X^\alpha Z^\beta Z^\delta X^\gamma \nonumber \\
&=&\frac{1}{2^{2n}}\sum_{\alpha,\beta}a_{\alpha,\beta}\sum_{\gamma,\delta}
(-1)^{\alpha\cdot\delta\oplus\gamma\cdot\beta}X^\alpha Z^\beta \nonumber \\
&=&\sum_{\alpha,\beta}a_{\alpha,\beta}
\delta_{\alpha,0}\delta_{\beta,0}X^\alpha Z^\beta \nonumber \\
&=&a_{0,0}I=\frac{Tr(\rho)}{2^n}I=\frac{1}{2^n}I
\end{eqnarray} 

\section{An Equivalent Problem}
Since there are a continuum of valid density matrices, the quantum security 
criterion (\ref{qcrypt_cond}) can be unwieldy to
deal with.  Here we introduce a modified condition that is
necessary and sufficient for security.
\begin{lm} \label{lemma_mod}
\em An encryption set $\{p_k,U_k\}$ satisfies eqn. 
(\ref{qcrypt_cond}) if and only if it satisfies:
\begin{equation} \label{mod_qcrypt_cond}
\sum_{k=1}^M p(k)U_k X^\alpha Z^\beta U_k^\dagger=\delta_{\alpha,0}\delta_{\beta,0}I\ .
\end{equation}
\end{lm}

{\bf Proof:} To show that the above condition
is sufficient, express $\rho$ in the $X^\alpha Z^\beta$ basis, as was
done in eqn.~(\ref{xz_proof}) and apply the eqn. (\ref{mod_qcrypt_cond}).
\begin{eqnarray*}
\sum_{k=1}^M p(k)U_k\rho  U_k^\dagger&=&\sum_{k=1}^M p(k)U_k
\left( \sum_{\alpha,\beta}a_{\alpha,\beta}X^\alpha Z^\beta \right)  
U_k^\dagger \\
&=&\sum_{\alpha,\beta}a_{\alpha,\beta}
\sum_{k=1}^M p(k)U_k X^\alpha Z^\beta U_k^\dagger \\
&=&\sum_{\alpha,\beta}a_{\alpha,\beta}\delta_{\alpha,0}\delta_{\beta,0}I \\
&=&a_{0,0}I=\frac{Tr(\rho)}{2^n}I=\frac{1}{2^n}I
\end{eqnarray*}

To show that the modified condition eqn. (\ref{mod_qcrypt_cond}), 
is necessary is  somewhat more involved.  First let us introduce some 
new notations:
\begin{eqnarray*}
\rho_i=\frac{I+\sigma_i}{2} \; &\;\hbox{and}\; \;& \rho_{mix}=\frac{I}{2} \ .
\end{eqnarray*}
The proof may be obtained by induction.  Suppose all $X^\alpha$ with 
$|\alpha|\le k$ are mapped to zero by the encryption process.  Now consider
the following product state  of $n-k-1$ mixed states, with exactly $k+1$ pure
states $\rho_x$:
\begin{eqnarray*}
\rho&=&\rho_{mix}\otimes\rho_{mix}\otimes\ldots\otimes\rho_{mix}
\otimes\rho_x\otimes\rho_x\otimes\ldots\otimes\rho_x
\end{eqnarray*}
By expanding the above becomes:
\begin{eqnarray*}
\rho&=&\frac{I}{2^n}+\frac{1}{2^n}\sum_{\alpha=1}^{2^k-1}X^\alpha
+\frac{1}{2^n}X^{2^{k+1}-1} 
\end{eqnarray*}
In the above we use decimal numbers where before we defined $X^\alpha$ with 
$\alpha$ in binary; hence $X^3=X^{00\ldots 011}$.
When the above $\rho$ is encrypted we know that $\frac{I}{2^n}$ is
mapped to itself.  By assumption $X^\alpha$ with $|\alpha|\le k$ is mapped
to zero, hence the sum in the expansion of $\rho$ disappears.  
Since $\rho$ must be mapped to $\frac{I}{2^n}$, then the
last term in the above, which is $X^\alpha$ with $|\alpha|=k+1$, must
be mapped to zero.  By permuting the initial input states, all
$X^\alpha$ with $|\alpha|=k+1$ must be mapped to zero.  The case where
$k=1$ is our base case.  By induction all $X^\alpha$ are mapped to zero.

If $x$ is replaced by $z$ in the above, then all $Z^\beta$ are mapped to zero
also.  If $x$ is replaced by $y$ and using the fact that all $X^\alpha$ and
$Z^\beta$ are mapped to zero, one sees that all $X^\alpha Z^\beta$ are mapped
to zero, which proves the lemma.\\
\mbox{} \hfill $\qed$

Thus, by using a basis for the set of $2^n\times2^n$ matrices, 
the condition for security becomes discrete, and only $2^{2n}$ 
equations need to be satisfied by the set $\{p_k, U_k\}$.  The above lemma
will be useful for showing necessary conditions on encryption sets.

\section{Characterization and Optimality of Quantum One-Time Pads}
So far, we have provided one quantum encryption protocol based on bit-wise Pauli rotations,
which uses $2n$ random classical bits in order to encrypt $n$ quantum bits.  
In this section we explore the following questions:
(1) What are some of the other choices of $\{p_k,U_k\}$ that can be used to perform quantum
encryption? In general, can one precisely characterize all possible valid choices of 
$\{p_k,U_k\}$? and (2) Is the simple quantum one time pad protocol optimal? That is,
can one encrypt $n$-bit quantum states using less than $2n$ random secret classical bits? 
First, we prove a sufficient condition for choosing a secure encryption 
protocol,  and then provide a corresponding necessary condition as well. In particular,
we show that one {\em cannot} perform secure encryption of $n$-bit quantum states using less
than $2n$ random classical bits.  

\begin{lm}\label{lemm_suff}\em 
Any unitary orthonormal basis for the $2^n\times2^n$ matrices
uniformly applied encrypts $n$ quantum bits. \label{suff_encryp_lm}
\end{lm} 
{\bf Proof}: We can always write the matrices, $U_k$, in terms of the $X^\alpha Z^\beta$ basis as
\begin{equation}
U_k=\sum_{\alpha,\beta}C_{\alpha,\beta}^k X^\alpha Z^\beta\ . 
\end{equation} 
Since these $U_k$'s form an orthonormal basis, the $2^{2n} \times 2^{2n}$ 
transformation matrix $C$, 
comprising of the transformation coefficients, is a unitary matrix. Hence, 
the rows and columns of $C$ are orthonormal:
\begin{equation} 
\sum_{k=1}^M C_{\alpha,\beta}^k (C_{\gamma,\delta}^k)^*=\delta_{\alpha,\gamma}
\delta_{\beta,\delta} \; \hbox{and} \; 
\sum_{\alpha,\beta} C_{\alpha,\beta}^k (C_{\alpha,\beta}^l)^*=\delta_{k,l}\ .
\end{equation}

By substitution of $U_k$ in (\ref{qcrypt_cond}) the lemma is obtained:
\begin{eqnarray*}
\frac{1}{2^{2n}}\sum_k U_k\rho U_k^\dagger&=&
\frac{1}{2^{2n}}\sum_k\left(\sum_{\alpha,\beta}C_{\alpha,\beta}^k 
X^\alpha Z^\beta\right)\rho\left(\sum_{\gamma,\delta}{C_{\gamma,\delta}^k}^*
Z^\delta X^\gamma\right) \\
&=&\frac{1}{2^{2n}}\sum_k\sum_{\alpha,\beta} 
\sum_{\gamma,\delta}C_{\alpha,\beta}^k{C_{\gamma,\delta}^k}^*
X^\alpha Z^\beta\rho Z^\delta X^\gamma \\
&=&\frac{1}{2^{2n}}\sum_{\alpha,\beta} 
\sum_{\gamma,\delta}\left(\sum_k C_{\alpha,\beta}^k
{C_{\gamma,\delta}^k}^*\right)
X^\alpha Z^\beta\rho Z^\delta X^\gamma \\
&=&\frac{1}{2^{2n}}\sum_{\alpha,\beta} 
\sum_{\gamma,\delta} \delta_{\alpha,\gamma}\delta_{\beta,\delta}
X^\alpha Z^\beta\rho Z^\delta X^\gamma \\
&=&\frac{1}{2^{2n}}\sum_{\alpha,\beta} 
X^\alpha Z^\beta\rho Z^\beta X^\alpha \\
&=&\frac{1}{2^n}I
\end{eqnarray*}
\mbox{} \hfill $\qed$

\begin{lm}\label{lemm_Cperp_C} \em
Given any quantum encryption set, $\{ p_k, U_k\}$, $k=1,\cdots,M$, (i.e.,$\displaystyle 
\sum_k p_k = 1$, $U_k$ is unitary, and eqns. (\ref{qcrypt_cond}) and (\ref{mod_qcrypt_cond}) 
are satisfied), let $\displaystyle \tilde{U_k}=\sqrt{p_k} U_k=\sum_{\alpha,\beta}
\tilde{C}_{\alpha,\beta}^k X^\alpha Z^\beta, $ and let $\tilde{C}$ be the $M\times 2^{2n}$ 
transformation matrix, comprising of the transformation coefficients $\tilde{C}_{\alpha,\beta}^k$. 
Then $M\ge 2^{2n}$, and
\[ \tilde{C}^\dag \tilde{C} = \frac{1}{2^{2n}} I_{2^{2n}\times 2^{2n}} \ .\]
\end{lm} 
{\bf Proof:} \hspace{1mm}  $\{ p_k, U_k\}$ satisfies eqns. (\ref{qcrypt_cond}) and 
(\ref{mod_qcrypt_cond}). Hence, for every $\ell, m$ $\in \{0,1\}^n$, 
\begin{eqnarray*}
\delta_{\ell,0}\delta_{m,0}I&=& \sum_{k=1}^M p(k)U_k X^\ell Z^m U_k^\dagger
\nonumber \\
&= &\sum_{k=1}^M \tilde{U_k} X^\ell Z^m \tilde{U_k}^\dagger \nonumber \\
&=& \sum_{k=1}^M\sum_{\alpha,\beta}\sum_{\gamma,\delta} \tilde{C}_{\alpha,\beta}^k 
(\tilde{C}_{\gamma,\delta}^k)^* X^\alpha Z^\beta X^\ell Z^m Z^\delta X^\gamma \nonumber \\
&=&\sum_{\alpha,\beta}\sum_{\gamma,\delta}(-1)^{\beta\cdot \ell+\gamma\cdot(\beta+\delta+m)} 
\left(\sum_{k=1}^M \tilde{C}_{\alpha,\beta}^k (\tilde{C}_{\gamma,\delta}^k)^* \right)
X^{\alpha+\gamma+\ell} Z^{\beta+\delta+m} \nonumber \\
&=&\sum_{p,q}\left(\sum_{\alpha,\beta} (-1)^{\beta\cdot \ell+(p+\ell+\alpha)\cdot q} 
\left(\sum_{k=1}^M \tilde{C}_{\alpha,\beta}^k (\tilde{C}_{\alpha+p+\ell,\beta+q+m}^k)^* \right)
 \right) X^p Z^q  .
\end{eqnarray*}
Using the linear independence of the $X^p Z^q$, only the identity component is non-zero.
Hence security implies:
\begin{eqnarray} \label{inner-prod-eqn}
\delta_{\ell,0}\delta_{m,0}\delta_{p,0}\delta_{q,0}&=&
\sum_{\alpha,\beta} (-1)^{\beta\cdot \ell+\alpha\cdot q}\left(
\sum_{k=1}^M \tilde{C}_{\alpha,\beta}^k (\tilde{C}_{\alpha+p+\ell,\beta+q+m}^k)^* \right) 
\nonumber \\
&=&\sum_{\alpha,\beta,\gamma,\delta} (-1)^{\beta\cdot \ell+\alpha\cdot q}
\delta_{\gamma,\alpha+p+\ell}\delta_{\delta,\beta+q+m}
\left(
\sum_{k=1}^M \tilde{C}_{\alpha,\beta}^k
(\tilde{C}_{\gamma,\delta}^k)^* \right)
\end{eqnarray}
As it will be evident, the second step in the above equation will be used 
to introduce a linear algebra formulation of the problem. Now, let
\begin{eqnarray*}
\Psi_{(\alpha,\beta),(\gamma,\delta)}&=& 
\sum_{k=1}^M \tilde{C}_{\alpha,\beta}^k (\tilde{C}_{\gamma,\delta}^k)^*\ ,
\end{eqnarray*}
which is the
standard inner product of the $(\alpha,\beta)^{th}$ and the $(\gamma,\delta)^{th}$
columns of $\tilde C$ or 
$\left(\tilde{C}^\dag \tilde{C}\right)_{(\alpha,\beta),(\gamma,\delta)}$, 
and let
\begin{eqnarray*}
{\bf M}_{(\ell,m,p,q),(\alpha,\beta,\gamma,\delta)}&=& (-1)^{\beta\cdot \ell+\alpha\cdot q}
\delta_{\gamma,\alpha+p+\ell}\delta_{\delta,\beta+q+m}\ .
\end{eqnarray*} 
Eqn. (\ref{inner-prod-eqn}) can now be written as a set of 
$2^{4n}$ linear equations: ${\bf M} {\bf \Psi}$ $=[1\; 0\; \cdots 0\ ]^T$, where ${\bf \Psi}$
is the $2^{4n} \times 1$ vector consisting of all the possible inner products of pairs of
columns of $\tilde{C}$, and $\bf M$ is a $2^{4n}\times 2^{4n}$ matrix with elements from
the set ${1, 0, -1}$.
Next we observe that a matrix $\bf A$ is orthogonal if and only if 
$\sum_j A_{i,j}A_{i',j}=A_i^2\delta_{i,i'}$, where $A_i$ is the norm of the $i^{th}$ row (which must be greater than zero).
One can easily verify that $\bf M$ is an orthogonal matrix:
\begin{eqnarray*}
&&\sum_{\alpha,\beta,\gamma,\delta}{\bf M}_{(\ell,m,p,q),(\alpha,\beta,\gamma,\delta)}
{\bf M}_{(\ell',m',p',q'),(\alpha,\beta,\gamma,\delta)}\\
&=&\sum_{\alpha,\beta,\gamma,\delta}(-1)^{\beta\cdot\ell+\alpha\cdot q}
\delta_{\gamma,\alpha+p+l}\delta_{\delta,\beta+q+m}
(-1)^{\beta\cdot\ell'+\alpha\cdot q'}
\delta_{\gamma,\alpha+p'+l'}\delta_{\delta,\beta+q'+m'} \\
&=&\sum_{\alpha,\beta,\gamma,\delta}(-1)^{\beta\cdot(\ell+\ell')+\alpha\cdot (q+q')}
\delta_{\gamma,\alpha+p+l}\delta_{\delta,\beta+q+m}
\delta_{\gamma,\alpha+p'+l'}\delta_{\delta,\beta+q'+m'} \\
&=&\sum_{\alpha,\beta}(-1)^{\beta\cdot(\ell+\ell')+\alpha\cdot (q+q')}
\delta_{p+l,p'+l'}\delta_{q+m,q'+m'}\\
&=&2^{2n}\delta_{l,l'}\delta_{q,q'}\delta_{p+l,p'+l'}\delta_{q+m,q'+m'}\\
&=&2^{2n}\delta_{l,l'}\delta_{q,q'}\delta_{p,p'}\delta_{m,m'}\ .
\end{eqnarray*} 
In showing the above we have also found the inverse of $\bf M$.  The
orthonormality of $\bf M$ means that ${\bf M}{\bf M}^T=2^{2n}I$, and hence
${\bf M}^{-1}={\bf M}^T/2^{2n}$.  Therefore,  
${\bf \Psi}$ $=\frac{{\bf M}^T[1\; 0\; \cdots 0\ ]^T}{2^{2n}}$, which means
${\bf \Psi}$ is the first row of ${\bf M}$ renormalized:
\[ \Psi_{(\alpha,\beta),(\gamma,\delta)} =
\frac{{\bf M}_{(0,0,0,0)(\alpha,\beta,\gamma,\delta)}}{2^{2n}}
=\frac{1}{2^{2n}} \delta_{\alpha,\gamma}
\delta_{\beta,\delta}\ . \]

Since $\left(\tilde{C}^\dag \tilde{C}\right)_{(\alpha,\beta),(\gamma,\delta)}=
\Psi_{(\alpha,\beta),(\gamma,\delta)}$ we have  
\[ \tilde{C}^\dag \tilde{C} = \frac{1}{2^{2n}} I_{2^{2n}\times 2^{2n}} \ . \]
Since $I_{2^{2n}\times 2^{2n}}$ is a full rank matrix, then $\tilde{C}$ must have at least
as many rows as columns.  $\tilde{C}$ has $2^{2n}$ columns so $M\ge 2^{2n}$.
\mbox{} \hfill $\qed$

\begin{thm}\em \label{opt_thm}
Any given quantum encryption set, $\{ p_k, U_k\}$, $k=1,\cdots,M$, 
(i.e.,$\displaystyle \sum_k p_k = 1$, $U_k$ is unitary, and eqns. 
(\ref{qcrypt_cond}) and (\ref{mod_qcrypt_cond}) are satisfied)
has:
\begin{eqnarray*}
H(p_1, \cdots, p_M)&=&\sum_{i=1}^M p_i \log \frac{1}{p_i}\geq 2n .
\end{eqnarray*}
Hence, {\it one must use at least $2n$ random classical bits for 
any quantum encryption}. 
Additionally, if $M=2^{2n}$, then $p_k=\frac{1}{2^{2n}}$ and $U_k$'s 
form an orthonormal basis. Hence, {\it a set $\{p_k,U_k\}$ involving only
$2n$ secret classical bits is a quantum encryption set if and only
if the unitary matrix elements form an orthonormal basis, and they 
are all equally likely}. 
\end{thm}

{\bf Proof:}
By Lemma \ref{lemm_Cperp_C} we have that
\[ \tilde{C}^\dag \tilde{C} = \frac{1}{2^{2n}} I_{2^{2n}\times 2^{2n}} \ . \]
Using a singular value decomposition\cite{GoV:89}
of $\tilde C$, we have the following relationships:
\[ \tilde{C}=W\Lambda V^\dagger,  \;\; \tilde{C}^\dag \tilde{C}= V (\Lambda^\dagger \Lambda) 
V^\dagger, \; \hbox{and} \; \tilde{C}\tilde{C}^\dag = W (\Lambda \Lambda^\dagger) W^\dagger\, ,
\]
where $W$ and $V$ are $M\times M$ and $2^{2n}\times 2^{2n}$ unitary matrices, respectively,
and $\Lambda$ is an $M\times 2^{2n}$ diagonal rectangular matrix: $\Lambda(i,j)= 
\lambda_i\delta_{i,j}$.Note that $\Lambda^\dagger \Lambda$
and $\Lambda \Lambda^\dagger$ are real diagonal matrices and have the same non-zero elements; 
hence, $\tilde{C}^\dag \tilde{C}$ and $\tilde{C}\tilde{C}^\dag$ have the {\em same} 
non-zero eigenvalues. 
Since $\tilde{C}^\dag \tilde{C}$ has $2^{2n}$ repeated eigenvalues 
($=\frac{1}{2^{2n}}$) and $M\geq 2^{2n}$,$\tilde{C}\tilde{C}^\dag$ 
has $2^{2n}$ repeated eigenvalues ($=\frac{1}{2^{2n}}$) and
the rest of its $M-2^{2n}$ eigenvalues are 0. Also note that the diagonal entries of 
$\tilde{C}\tilde{C}^\dag$ are the probabilities $p_k$'s and hence,
\[ p_k =\frac{Tr(\tilde{U}_k{\tilde{U}_k}^\dagger)}{2^n}=(\tilde{C}\tilde{C}^\dag)_{k,k}= \frac{1}{2^{2n}} \sum_{i=1}^{2^{2n}}
|W_{i,k}|^2 \leq 
\frac{1}{2^{2n}}\ . \]
The above uses the facts that since $W$ is unitary, $\sum_{i=1}^{M}|W_{i,k}|^2=1$ and 
that $M\ge 2^{2n}$.
Hence, 
\[ H(p_1, \cdots, p_M) = \sum_{i=1}^M p_i \log \frac{1}{p_i}\geq 2n \sum_{i=1}^M p_i = 2n\ . \]
In the particular case where $M=2^{2n}$, we have $\displaystyle \tilde{C}\tilde{C}^\dag= 
\tilde{C}^\dag \tilde{C} = \frac{1}{2^{2n}} I_{2^{2n}\times 2^{2n}}$.
Hence
\begin{eqnarray*}
\frac{Tr(\tilde{U}_k{\tilde{U}_j}^\dagger)}{2^n}&=&
\delta_{k,j}\frac{1}{2^{2n}} \ ,
\end{eqnarray*}
which gives
$p_k=\frac{1}{2^{2n}}$,
and that the set $\{ U_k\}$ necessarily forms an orthonormal basis.
The proof is completed by observing that by lemma \ref{lemm_suff} any 
unitary orthonormal basis applied uniformly is sufficient.
\mbox{} \hfill $\qed$

\section{Encryption vs. Teleportation and Superdense Coding}
One of the most interesting results in quantum information theory
is the teleportation of quantum bits by shared EPR pairs and classical
channels\cite{teleport}.  The quantum one time pad described in 
Section~\ref{onetime_sec} could be implemented using the usual 
teleportation scheme by encrypting the classical communications 
with a one time pad.  Hence, teleportation gives one example of 
a quantum encryption algorithm. In the original teleportation 
paper\cite{teleport} a proof that
two classical bits are required to teleport is given.  The proof
is based on a construction that gives superluminal communication
if teleportation can be done with less than two bits.  This proof
however does not imply that all quantum encryption sets require $2n$
bits.  To do so would require one to prove that all quantum encryption
sets correspond to a teleportation protocol.
On the other hand, as we show next, all teleportation protocols correspond to
a quantum encryption set; hence, {\it Theorem~\ref{opt_thm} provides a
new proof of optimality of teleportation}.  

A general teleportation scheme can be described as follows: Alice and Bob share
a pure state comprising $2n$ qubits, $\rho_{AB}$, such 
that the traced out $n$-bit states of Alice and Bob satisfy: $\rho_{A} = \rho_{B}
=\frac{1}{2^n} I$. Next, Alice receives an unknown $n$-bit quantum state $\rho$, and performs
a joint measurement (i.e., on $\rho$ and $\rho_A$), which
produces one of a fixed set of outcomes $m_k$, $k=1,\ldots,M$, each with probability $p_k$. The
particular outcome $m_k$ is classically communicated to Bob using $H(p_1, \ldots, p_M)$ bits. Bob
performs a corresponding unitary operation $U_k$ on his state to retrieve $\rho$. Hence, 
after Alice's measurement (and before Bob learns the outcome), Bob's state can be expressed as
$\rho_B=\frac{1}{2^n} I=\displaystyle \sum_{k=1}^M p(k) U_k\rho U_k^\dag$, which is exactly
the encrypted state of the message, $\rho$, defined in Eqn. (\ref{qcrypt_cond}). Hence,
every teleportation scheme corresponds to an encryption protocol $\{ p_k, U_k\}$. Since we prove
that all quantum encryption sets require $2n$ classical bits, 
then all teleportation schemes must also require $2n$ classical bits.
Note that our proof only relies on the properties of the underlying
vector spaces.

Superdense coding\cite{superdense} also has a connection to quantum
encryption.  Consider the case where Alice asks Bob to encrypt
something and then Alice wishes to learn the key that Bob used
to encrypt.  In the case of the classical one time pad \cite{one_time_pad}
$c=m\oplus k$, and so given a message and it's accompanying ciphertext,
one learns the key: $k=m\oplus c$.  Quantumly, each quantum bit
has two classical key bits to learn.  Due to Holevo's
theorem\cite{Holevo73}
it may seem that this implies that there is no way to learn the classical
key exactly.  This intuition is not correct.  
Alice can learn Bob's key in the following way.
Alice prepares $n$ singlets
and gives half of each singlet to Bob.  Bob encrypts them using the simple
quantum one time pad and returns them to Alice. Alice can learn the
key exactly by measuring each former singlet in the bell basis.  The
outcome would tell Alice exactly which transformation Bob applied.
This protocol corresponds exactly to the superdense coding scheme\cite{superdense}.

Interestingly, some insight is gained as to where the factor of 
two between the number of classical and quantum bits comes from in 
both encryption and teleportation.  In the case 
of classical bits, $\rho$ is diagonal.  A basis for all diagonal 
matrices is $Z^\beta$.  Hence, 
for encryption of classical bits there are only $2^n$ equations.
In the quantum case, by lemma \ref{lemma_mod}, there are $2^{2n}$
equations to satisfy, so it is not too surprising that
there are twice as many classical bits needed.  Equivalently,
the $\log$ of the size of the space is twice as large quantumly
as opposed to classically.
The {\em proof given here could be particularized to give a new
proof of Shannon's original result} on informationally secure
classical encryption\cite{one_time_pad}.

\section{Discussion}
We have presented an algorithm for using $2n$ secret classical
bits to secure $n$ quantum bits.  These encrypted quantum 
bits may now be held by an untrusted party with no danger that 
information may be learned from these bits.  
Any number of applications may be imagined for
this algorithm, or class of algorithms $\{p_k,U_k\}$.  For instance,
rather than using random classical data of size $2n$, one could
use a secret key ciphers\cite{Schneier:AC} or stream ciphers\cite{Schneier:AC}
to keep a small finite classical key, for instance 256 bits,
to generate pseudo-random bits to encrypt quantum data.  In fact,
these notions allow for straight-forward
generalizations of many classical protocols to quantum data.  Quantum
secret sharing has been developed\cite{quantum_secret_sharing} that may
be used to share quantum secrets.  Classical secret sharing schemes
are known that are informationally secure\cite{class_secret_sharing}.
By encrypting a quantum state of $n$ bits with $2n$ classical bits,
and then using classical secret sharing on the $2n$ bits, one
may use these informationally secure classical methods in the quantum
world.  This protocol would allow users with only classical resources
to perform secret sharing given an untrusted center to store
the quantum data. One application independently suggested by 
Cr\'epeau et. al.\cite{CDM_NEC} is to build quantum bit commitment 
schemes based on computationally secure classical bit commitment schemes.

\section{Acknowledgements}
We would like to thank Tal Mor for helpful discussions.
This work 
was supported in part by grants from the Revolutionary Computing
group at JPL (contract \#961360), and from the DARPA Ultra program
(subcontract from Purdue University \#530--1415--01)

\end{document}